\documentclass[preprint,showpacs,preprintnumbers,amsmath,amssymb]{revtex4}

\usepackage{graphicx}% Include figure files
\usepackage{bm}
\usepackage{multirow}
\usepackage{verbatim}

\newcommand{\acp}{\ensuremath{{\cal A}_{CP}\ }}

\newcommand{\ButoKpipi} {\ensuremath{ B^\pm \to K^\pm \pi^+ \pi^- }}
\newcommand{\BdtoKpipi} {\ensuremath{ B^0,\bar{B^0} \to K_S \pi^+ \pi^- }}

\begin{document}
\title{ Extracting CKM $\gamma$ phase from $B^{\pm} \to K^{\pm} \pi^+ \pi^-$ 
and \\
 $B^0$,   $\bar B^0 \to K_s \pi^+ \pi^-$ }

\author{Ignacio Bediaga, Gabriel Guerrer and Jussara M. de Miranda \\  
	bediaga@cbpf.br, guerrer@cbpf.br and jussara@cbpf.br \\ $\,$ \\}
  
\affiliation{Centro Brasileiro de Pesquisas F\'\i sicas, Rua Xavier Sigaud 150,
 22290-180  -- Rio de Janeiro, RJ, Brazil\\}

%%%%%%%%%%%%%%%%%%%%%%%%%%%%%%%%%%%%%%%%%%%%%%%%%%%%%%%%%%%%%%%%%%%%%%%%%%%%%%%%%
\begin{abstract}
We discuss some aspects of the search for CP asymmetry in the three body B decays,
revealed through the interference 
among  neighbor resonances in the Dalitz plot.
We propose  a competitive  method to extract 
the CKM  $\gamma$ angle 
combining Dalitz plot amplitude  analysis of $B^{\pm} \to K^{\pm} \pi^+ \pi^-$ 
and untagged $B^0$, $\bar B^0 \to K_s \pi^+ \pi^-$. 
The method also  obtains the ratio 
and phase difference between the {\it tree} and  {\it penguin} contributions 
from  $B^0$ and  $\bar B^0 \to K^{*\pm} \pi^{\mp} $ decays and the CP 
asymmetry between $B^0$ and  $\bar{B^0}$. 
From Monte Carlo studies 
of 100K events for the neutral mesons, we show the possibility of 
measuring $\gamma$.
\end{abstract}

\maketitle

%%%%%%%%%%%%%%%%%%%%%%%%%%%%%%%%%%%%%%%%%%%%%%%%%%%%%%%%%%%%%%%%%%%%%%%%%%%%%%%%%%%%%%%%%%%%%%%%%%%%%%%%%%%%%%%
\section{Introduction}

 According to the  Standard Model, CP violation is only possible 
 through the CKM complex  parameters. Interference between processes with 
 different weak phases  contributing to a same final state  can generate 
 an asymmetry in charge conjugate $ B $ meson decays. 
 The most  established methods to extract the CKM  phases  
 exploits  the  interference between the phases generated in the
 $B^0$  $\bar B^0$ oscillation and the  phase of the decay of 
 these particles in the same final state.  Using such approaches  on the 
 $B^0 \to J/\Psi K_s$ decay, the CKM $\beta$  angle is measured to 
 a good precision. However,  the other two CKM angles, 
 $\alpha$ and $\gamma$, retain  sizable uncertainties, 
  due to both experimental  and theoretical difficulties. LHC   
 should  determine $\gamma$    to $\sim 5^0 $ in one year of data taking\cite{schneider}. 
 
 Any asymmetry observed in charge conjugated $B$ decays is interpreted as 
 a manifestation of the CKM phases. In the 
 two body decay  scenario,  direct $CP$ asymmetry (\acp) was established by 
 the experiments 
 BaBar \cite{BaBarkpi} and Belle \cite{bellekpi} by simply counting 
 the difference in number of  events of the  $B^0 \to K^+\pi^-$ and
  $\bar B^0 \to K^-\pi^+$ decays. 
 In three body $B$ meson decays,  BaBar and Belle 
  \cite{BaBark2pi,  bellek2pi}  claim the observation of direct \acp in the channel  $B^{\pm} \to K^{\pm} \rho^0$ measured through 
   the amplitude analysis of the final state $B^{\pm} \to K^{\pm} \pi^+ \pi^-$.
     They see a difference  in  the decay fractions 
     among the intermediate states in the $B^+$ and $B^-$ samples.  That is  
     the difference between the square of the fitted  amplitude 
     for the  intermediate  state integrated on the phase space. 
     This quantity is a straightforward  application of the two body 
     strategy into the  three body decays,  including the requirement of a 
     strong phase
      difference to observe  \acp.

 Beyond  the  fraction  asymmetry quoted by Belle and BaBar, one
 could explore the  asymmetry  associated with the interference among 
 two intermediate  states, i.e. the interference between two neighbor 
 resonances decaying to the same three body final state. 
As will be discussed in detail in section III, whenever the phase difference 
between two   resonant amplitudes involves the  weak phase, one should expect 
 to see \acp in the interference terms of the Dalitz plot  
 distribution. This contribution to the
 asymmetry  depends only on the existence  of a weak
  phase difference and it can be observed even in situations where there is 
  no strong phase  difference. There are a lot of opportunities for
  investigation of the asymmetry in the Dalitz plot distribution 
   rather than difference  in  the decay fractions alone\cite{burdman,asner}.

The idea of extracting $\gamma$ exploring the Dalitz interference  among   
intermediate  amplitudes in three body $B$ decays   was initially proposed 
for $ B^{\pm} \to \pi^{\pm} \pi^+ \pi^-$ \cite{bbgg},
where $\chi_{c0}$ plays a fundamental role as reference channel.  
However, the method is statistically limited by the low contribution of the
Cabibbo suppressed amplitude  $ B^{\pm} \to \chi_{c0} \pi^{\pm}$ \cite{BaBar3pi}, 
 and also estimated through the already observed Cabibbo allowed 
$B^{\pm}\to \chi_{c0} K^{\pm}$ 
channel \cite{BaBark2pi,  bellek2pi}. 
Recently, a new effort has been  made to study
$\gamma$ in $B\to K \pi\pi$ decays, however this  approach requires 
time-dependent analysis \cite{cps, gpsz}, 
 having to deal with tagging inefficiencies that could be as large as 
 90\% at LHCb \cite{TDRLHCb}.

We present a new possibility to measure the 
 CKM $\gamma$ angle,  through the  interference terms in  the amplitude 
 analysis of three body $B$ decays. The method combines  the experimental 
 data from    $B^{\pm} \to K^{\pm} \pi^+ \pi^-$ and  $B^0$,  
  $\bar B^0 \to K_s \pi^+ \pi^-$. From the charged $B $  decays 
 we have the {\it penguin}  amplitude coefficient and phase of the channel 
 $K^* \pi$. We then use this result in an untagged analysis of the neutral
 system, to extract the {\it tree } component strong 
 phase and amplitude coefficient and  the weak phase  $\gamma$. The method is 
 based on the ability to measure independently the relative amplitudes and 
 phases for $B^0$ and $\bar B^0$ decays in a joint  untagged sample.

%%%%%%%%%%%%%%%%%%%%%%%%%%%%%%%%%%%%%%%%%%%%%%%%%%%%%%%%%%%%%%%%%%%%%%%%%%%%%%%%%%%%%%%%%%%%%%%%%%%%%%%%%%%%%%%
\section{Amplitude Analysis}

The standard  amplitude analysis, extensively used in
 non-leptonic three body heavy quark meson  decays,  is based on the Isobaric model.
 In these analyses, the two body resonant intermediate  state amplitudes are
 represented by Breit-Wigner functions multiplied by angular distributions
 associated with the spin of the resonance.
 The various contributions are combined in a coherent sum with complex
 coefficients that are  extracted from maximum likelihood fits to the data.
 The absolute value of the coefficients, normally refereed to as magnitudes,
  are related to the relative fraction of each intermediate  state  and the phase
   takes into account rescattering or the final state interaction
  (FSI) between the resonance and the bachelor particle. This phase is
  considered constant because it depends only on the total energy of the
  system, i.e. the heavy meson mass. The presence of strong final state
  constant  phases can be directly revealed through the interference
  between two different amplitudes. Thus, the total  Isobaric amplitude, written
  as functions of the Dalitz plot variables, $s_{12}$ and $s_{13}$ is:

\begin{equation}
{\cal A_T} = a_{NR} e^{i\delta_{NR}}{\cal A}_{NR}(s_{12},s_{13}) + 
\sum_{n=1}^N a_n e^{i\delta_j}{\cal A}_n(s_{12},s_{13}) \, .
\end{equation}
\noindent  
 
 The fit parameters are the  coefficient magnitudes, 
$a$, and the phases, $\delta$. The non-resonant amplitude, 
$ {\cal A}_{NR}  $ is usually represented by a constant. Each resonant amplitude,
 ${\cal A}_n( n \ge 1) $, is written as a product of three terms:

\begin{equation}
{\cal A}_n = \ ^{J}F_n \times {^J}{\cal M}_n \times BW_n
\end{equation}

\noindent The  first term are the Blatt-Weisskopf damping 
factors \cite{blatt}. 
$^{J}{\cal M}_n$ is a term which accounts for angular-momentum 
conservation and depends on the spin, $J$, of $n ^{th}$ the resonance, 
such that it is 
1 for a spin-0,  
${ -2\mid {\bf p_3} \mid \mid {\bf p_2} \mid cos\theta}$ for
spin-1, and ${ \frac{4}{3}(\mid {\bf p_3} \mid \mid {\bf p_2} \mid)^2 
(3cos^2\theta-1)}$ for spin-2, in the case of the resonance formed by the pair
of particles {\bf 1} and {\bf 3}.  ${\bf p_2}$  and ${\bf p_3}$  are the 
 the 3-momentum of particles {\bf 2}  and {\bf 3}, and $\theta$ is 
 the angle between particles {\bf 2} and {\bf 3},
  all measured in the resonance rest frame.
The last term is a relativistic Breit-Wigner function given by:

\begin{equation}
 BW_n = {\frac {m_n \Gamma_n} {m^2_n - s_{ij} - im_n\Gamma_n(s_{ij})}}, 
\end{equation}

\noindent where $m_n$ is the resonance mass and $\Gamma_n(s_{ij})$ is the mass 
dependent width.

The relative contribution of each intermediate  state is given by fractions defined as:

\begin{equation}
f_n \equiv {\frac {\int ds_{12} ds_{13} \mid c_n {\cal A}_n \mid^2}{ {\int
ds_{12}ds_{13}
\sum_{jk}  \mid c_j {\cal A}_j c_k^* {\cal A}_k^* \mid}}}\ .
\end{equation}

\noindent  where $\int ds_{12}ds_{13}\sum_{jk}  \mid c_j{\cal A}_j c_k^* {\cal A}_k^*  \mid$
is related to the total number of  events of the non-leptonic three body 
heavy meson  decay.

%%%%%%%%%%%%%%%%%%%%%%%%%%%%%%%%%%%%%%%%%%%%%%%%%%%%%%%%%%%%%%%%%%%%%%%%%%%%%5
 \section{The Direct Dalitz plot Asymmetry -${\cal A}_{DP}$ }

One important feature of the Dalitz plot associated with the interference
between two neighboring resonances is that there will be an    
\acp manifestation, independent of the strong phase differences.
To illustrate this point, let us develop a simple example of two  resonant 
scalar amplitudes, interfering  
in a particular region at the three body decay phase space. Suppose  one of the resonant amplitudes, $\mathcal{A}_1$, has a {\em tree} component,
with magnitude   {\bf $a_T$} and  strong phase {\bf $\delta_T$} and 
 weak phase  $\gamma$, plus a {\em penguin}  contribution with {\bf $a_P$} and 
 {\bf $\delta_P$} and no weak phase,  interfering with  another resonant 
amplitude, $\mathcal{A}_R$, with magnitude and phase {\bf $a_R$} and 
{\bf $\delta_R$} and the  absence of weak phase. The  charge conjugate total 
decay amplitudes are:

\begin{eqnarray}
A^+ = [ a_T e^{i (\delta_T + \gamma)} + a_P e^{i \delta_P } ]  
\mathcal{A}_1 + a_R e^{i \delta_R } \mathcal{A}_R \\
A^- = [ a_T e^{i (\delta_T - \gamma)} + a_P e^{i \delta_P } ]  
\mathcal{A}_1 + a_R e^{i \delta_R } \mathcal{A}_R 
\end{eqnarray}

and

\begin{eqnarray}
|A^+|^2 - |A^-|^2 = -4 a_T a_P \sin{\gamma} \sin{(\delta_T -  
\delta_P)} |\mathcal{A}_1|^2 \nonumber\\
-4 a_R a_T  \sin{\gamma} 
(\sin{(\delta_T - \delta_R)} \; {\mathcal  
Re} (\mathcal{A}_1 \mathcal{A}_2^*) 
+ \cos{(\delta_T - \delta_R)} \; {\mathcal  
Im} (\mathcal{A}_1 \mathcal{A}_2^*))
\end{eqnarray}

From Equation 7, we 
 see that the  first term has a  similar
behavior as the two body decays, i.e. \acp depends on the magnitude 
of the {\em tree} component and also on the phase difference $(\delta_T^1 -
\delta_P^1)$, which must be non zero for \acp to be observable. 
This is the term responsible for the  \acp quoted  through the
difference in integrated fractions made by Babar and Belle \cite{BaBarkpi} 
\cite{bellekpi}.
However the presence of  second or third terms, the interference ones,
 depends only on the 
magnitude of the {\em tree} component, since the phase difference has a 
 sine form associated with the  real part of the BW's product, but also a 
 cosine component associated with the imaginary part of the  BW's product.  
So the interference will  always produce  \acp provided 
that the magnitude associated with the weak phase is significant.

 Interferences in the Dalitz plot do not allow to associate directly
the fraction, defined in Eq. 4, with the number of events in a intermediate  
state channel.  The relative phase between two amplitudes sharing the same 
phase space region can produce destructive or constructive interferences, 
decreasing or increasing the number of events in the three body final sample, 
in which case we lose deterministic information of the source of the events in
the region. Due this difficulty to the concept of the direct \acp   to the three body decays,
 the asymmetry observed from  fractions would be better refered as 
{\it quasi-direct} \acp.

The direct \acp in the three body decays, in the sense of event 
counting,  would  be more conveniently defined by
 the notion of  two asymmetries, one local and one global. A global 
 \acp is the  difference between the total number 
 of events of two conjugate  three body decays. The  local would be the 
 difference in the number of events  
  depending on the Dalitz variables 
  $s_{ij}$. In general,  the  global asymmetry loses part of  
  dynamics information, since it is
   the integral over the phase space. On the other hand, local  asymmetry
    accounts for the difference of the magnitude square,
 or  the first term of Equation 7, and also  the asymmetry  coming from the 
   interference terms among two  resonances, or the last terms of
    Equation 7.

 The local CP asymmetry of  the charged B meson three body 
 decay   can be  defined through   the subtraction of the two Dalitz 
 surfaces relative to the charged conjugated B signals:

\begin{equation}
{\cal A}_{DP} = { \frac { N^+ (s_{12},s_{13}) -  N^- (s_{12},s_{13})}
 {N^+(s_{12},s_{13}) +  N^- (s_{12},s_{13})} }
\end{equation}

\noindent we call ${\cal A}_{DP}$  the local asymmetry at the Dalitz plot,
with $N^{\pm} (s_{12},s_{13})$ being the number of events in  
$s_{12}$, $ s_{13}$.

For the neutral systems with $B^0$ and 
$ \bar{B^0}$  decaying to the same final state,  $ K_S \pi^+ \pi^-$,  
 $ K_S p \bar p$,  $ \pi^0 \pi^+ \pi^-$ for example, Burdman and Donoghue \cite{burdman} 
 propose a convenient definition of local ${\cal A}_{DP}$. They showed that
 for a decay with $|p/q|=1$, the asymmetry is time independent and 
 can be written as:

\begin{equation}
{\cal A}_{dp} = { \frac { N^0 (s_{12},s_{13}) -  N^0 (s_{13},s_{12})}
 {N^0(s_{12},s_{13}) +  N^0 (s_{13},s_{12})} }
\end{equation}

A measure of local ${\cal A}_{DP}$  is obtained by 
comparing the  number of events in points of the Dalitz plot
symmetrical with respect to the diagonal   line $s_{12}=s_{13}$.
 
To quantify the total amount of asymmetry present in a three body decay,
one can compute the sum of the modulus of ${\cal A}_{DP}$, taking in account
properly statistical fluctuations.

%%%%%%%%%%%%%%%%%%%%%%%%%%%%%%%%%%%%%%%%%%%%%%%%%%%%%%%%%%%%%%%%%%%%%%%%%%%%%%%%%%%%%%%%%%%%%%%%%%%%%%%%%%%%%%%

\section{Method}
 
We now introduce a new possibility to measure $\gamma$, based
on a time-independent analysis of $B \to K \pi\pi$ decays. 
The most relevant observed intermediate states  for \ButoKpipi and \BdtoKpipi 
\cite{BaBark2pi, bellek2pi, bellekspipi}, are summarised in Table 
\ref{tab:bkpipicontrib} with the respective {\it tree} and {\it penguin} 
contributions, and CKM weak phases
contained in the amplitude. 
%Although $\gamma$ is present in some tree components, it's always followed by a 
%penguin component and also a phase from the strong final state interaction. 
%Since magnitudes and phases
%extracted by the fit procedure are the overall values, it is impossible to isolate and measure 
%$\gamma$ in a simple amplitude analysis of either charged or neutral $B$ meson decay.

{\small
\begin{table}[!h]
  \begin{center}
\begin{tabular}{c|c|c|c}
\centering
 & resonance & contribution & weak phase\\
\hline \hline
\multirow{5}{*}{$B^+ \to$} & $K^*(890)^0 \pi^+$ & \multirow{2}{*}{$V_{bt}V^*_{ts} \; P$} &
\multirow{2}{*}{} \\ \cline{2-2}
& $K^*(1430)^0 \pi^+$   & \\ \cline{2-4}
& $K^+ \rho(770)^0$ & \multirow{2}{*}{$V_{bt}V^*_{ts} \; P + V_{bu}V^*_{us} \; T^C_S$} &
\multirow{2}{*}{$\gamma$} \\ \cline{2-2}
& $K^+ f_0(980)$ & & \\ \cline{2-4}
& $K^+ \chi_{c0}$ & $V_{bc}V^*_{cs} \; T_S$ &   \\ \hline \hline
\multirow{5}{*}{$B^0 \to$} & $K^*(890)^+ \pi^-$ & \multirow{2}{*}{$V_{bt}V^*_{ts} \; P + V_{bu}V^*_{us} \; T^C$} &
  \multirow{2}{*}{$\gamma$} \\ \cline{2-2}
& $K^*(1430)^+ \pi^-$ &  &  \\ \cline{2-4}
& $K_S \, \rho(770)^0$ & \multirow{2}{*}{$V_{bt}V^*_{ts} \; P + V_{bu}V^*_{us} \; T_S$} &
 \multirow{2}{*}{$\gamma$} \\ \cline{2-2}
& $K_S \, f_0(980)$ & & \\ \cline{2-4}
& $K_S \, \chi_{c0}$ & $V_{bc}V^*_{cs} \; T_S$  \\
\hline \hline
\end{tabular}
\end{center}
	\caption{Resonances and dominant contributions from $B \to K \pi \pi$ decays . We denote
    allowed, suppressed by color and {\em penguin} amplitudes by $T^C$, $T_S$ and $P$.}
    \label{tab:bkpipicontrib}
\end{table}
}

\begin{comment}
Observing the penguin amplitudes of $B \to K^* \pi$,
Fig. \ref{fig:diagcomp}, we see that the difference between $B^-$ and $\bar{B^0}$ 
is on the quark permutation 
$u \leftrightarrow d$. According to SU(3) flavor symmetry, this permutation is
 a good symmetry given the close
mass values of quarks $u$ and $d$. In this case, we expect the same value in 
the contribution $a_P \, e^{i \, \delta_P}$
for the four processes $B^\mp \to K^* \pi^\mp$ and $\bar{B^0},B^0\to K^{*\mp} \pi^\pm$.
\begin{figure}[!htb]
\centering
	\includegraphics[scale=.4]{diagcomp}
	\caption{$B^- \to K^* \pi^-$ and $\bar{B^0}\to K^{*-} \pi^+$ diagrams, 
	with the same contribution 
	$a_P \, e^{i \, \delta_P}$
    according to flavor SU(3) symmetry.}
	\label{fig:diagcomp}
\end{figure}

\end{comment}

Based on SU(2) flavor symmetry, we expect the same {\em penguin} amplitudes
for the four processes $B^\pm \to K^* \pi^\pm$ and $B^0,\bar{B^0} \to K^{*\pm} \pi^\mp$.
This property allows to  extract the $B^\pm \to K^* \pi^\pm$ {\em penguin}
 parameters
{\footnote{For simplicity we refer only to $K^*$, while in practical the method uses 
simultaneously the parameters of the
resonances $K^*(890)$ and $K^*(1430)$.}}, and use them in the 
$B^0$ and $\bar{B^0}$ 
 joint fit to measure the {\em tree} phases, from which $\gamma$ is obtained.

%ancora
Since  we measure only relative  magnitudes and phases we must elect one resonance
to have fixed parameters. The compatibility
between {\em penguin} parameters from $B^\pm$ and $B^0,\bar{B^0}$ to $K^* \pi$, requires
that the amplitude analysis be made relative to an anchor resonance that has the same amplitude for $B$ charged and
neutral. For that we chose the non CP violating $B \to K \, \chi_{c0}$ amplitude.
The asymmetry measured by Belle \cite{bellechi} for the channel $B^\pm \to K^\pm \chi_{c1}$ is 
$A_{CP}=-0.01 \pm 0.03 \pm 0.02$, indicating 
that the dominant contribution is a {\em tree} diagram without weak phase.

Following  is a schematic representation of the method, with the arrows pointing its flow:

\begin{flushleft}
\begin{eqnarray}
B^\pm \to K^{*0} \pi^\pm \quad  : \quad {\underline a_P} \, e^{i \, {\underline \delta_P}} \hspace{1,7cm}	\nonumber \\
\Downarrow \hspace{2,1cm} \nonumber \\
B^0 \to K^{*+} \pi^-   \quad  \,  : \quad  a_P \, e^{i \, \delta_P} + {\underline a_T}
\, e^{i \, {\underline  \theta^+}} \nonumber \\
\bar{B^0} \to K^{*-} \pi^+  \quad : \quad  a_P \, e^{i \, \delta_P} + {\underline a_T} \, e^{i
\, {\underline \theta^-}} \nonumber \\
\Downarrow \hspace{0,5cm} \nonumber \\
\gamma = \frac{\theta^+ - \theta^-}{2}.
\label{eq:esquemag}
\end{eqnarray}
\end{flushleft}
 
 \noindent where $\theta^\pm=(\delta_T \pm \gamma)$. We underlined the parameters measured  in each
 amplitude analysis.

%hipoteses
The method is based on three basic and well accepted hypothesis that can be 
tested during the analysis. {\em Hypothesis 1}: the dominant contribution of 
$B^\pm \to K^* \pi^\pm$ is $V_{bt}V^*_{ts} \; P$. This has been
 indicated  by the BaBar result \cite{BaBark2pi}.
The test is to check if the $B^+ \to K^* \pi^+$ and $B^- \to K^* \pi^-$ 
amplitudes are the same. {\em Hypothesis 2}: the {\em penguin} components from 
$B^\pm \to K^* \pi^\pm$ and $B^0, \bar{B^0} \to K^{*\pm} \pi^\mp$
are equal \cite{teo-chiang, teo-beneke}.
%that explores the consequences of flavor SU(3) symmetry in the decay amplitudes.
{\em Hypothesis 3}: $\chi_{c0}$ have the same amplitude for $B^\pm \to K^\pm \chi_{c0}$ and
$B^0, \bar{B^0} \to K_S \, \chi_{c0}$. 
The experimental test for the second and third hypothesis, consists in the equality of the 
{\em tree} magnitudes measured 
from the intermediate  process $B^0, \bar{B^0} \to K^{*\pm} \pi^{\mp}$.
The confidence in the result of $\gamma$ is based in the fulfillment of all 
hypotheses,
whereas the failure of either implies  interesting unexpected effects,
like a big W annhilation contribution or even a new contribution for 
the $B^\pm \to K^\pm \chi_{c0}$ other than the tree component.

\section {$B^{\pm} \to K^{\pm} \pi^+ \pi^-$ 
and  $B^0$,   $\bar B^0 \to K_s \pi^+ \pi^-$ Amplitudes}

To obtain  the $B^\pm \to K^\pm \pi^+ \pi^-$  parameters, we apply the usual 
 maximum likelihood fit to the isobaric amplitudes given by Equation 2
 with the intermediate  states  $i=K^*(890)^0,\, K^*(1430)^0,\, 
 \rho(770)^0,\, f_0(980)$ and $K^+ \chi_{c0}$.

Regarding the neutral system, separation between $B^0$ and $\bar{B^0}$ to 
$K_S \pi^+ \pi^-$ is not trivial. The samples could be distinguished  at the
production level only through  partner $b$ particle tagging techniques. 
Furthermore, there is mixing among $B^0$ and $\bar{B^0}$, introducing time 
dependence in the decay probabilities.

The probability distribution for the final state $K_S \pi^+ \pi^-$, 
independent of its origin, is given by the incoherent sum: 
$|M(\Delta t)|^2+|\bar{M}(\Delta t)|^2$. The matrix element 
for observing the original $B^0$ and $\bar{B^0}$ decay to the common
final state are respectively:

\begin{eqnarray}
\! M(\Delta t)=e^{-(\Gamma /2 - i M) \Delta t} \, [ \mathcal{A} \, \cos{(\Delta m \, \Delta t/2)}  
 -i \, q/p \, \bar{\mathcal{A}} \, \sin{(\Delta m\, \Delta t/2)} ], \\ 
\bar{M}(\Delta t)= e^{-(\Gamma /2 - i M) \Delta t} \, [\bar{\mathcal{A}} \, \cos{(\Delta m \, \Delta t/2)} 
-i \, p/q \, \mathcal{A} \, \sin{(\Delta m\, \Delta t/2)}],
\end{eqnarray}

\noindent where $\mathcal{A}$ e $\bar{\mathcal{A}}$ are time-independent decay 
amplitudes equivalent to Equation 1, but
for $B^0 \to K_S \pi^+ \pi^-$ and $\bar{B^0} \to K_S \pi^+ \pi^-$ decays.

Assuming the same production rate for $B^0$ and $\bar{B^0}$, and in the case of $|p/q|=1$, it
was shown \cite{burdman, gardner}
 that this sum displays
the interesting property of canceling mixing  dependence terms,
\begin{equation}
|M(\Delta t)|^2+|\bar{M}(\Delta t)|^2 = e^{-\Gamma \, t} \, ( \, 
|\mathcal{A}^0|^2+|\bar{\mathcal{A}}^0|^2 \, ).
\label{eq:somaamps}
\end{equation}

 Similarly  to Equation 1, ${\mathcal{A}}^0$ and $\bar{\mathcal{A}}^0$ are
 written as: 

\begin{equation}
\mathcal{A}^0= a_\chi e^{i \, \delta_\chi} \, \mathcal{A}_\chi + \sum_i \, 
a_i e^{i \, \delta_i} \, \mathcal{A}_i
\end{equation}
\begin{equation} 
\bar{\mathcal{A} ^0} = \, \bar{a_\chi} e^{i \, \bar{\delta_\chi} } \, 
\mathcal{A}_\chi + \sum_i \, \bar{a_i} e^{i \, \bar{\delta_i} } 
\, \mathcal{A}_i
\end{equation}

As a fundamental step to achieve a $\gamma$ measurement, we propose a 
 method of \emph{joint fitting} the set of untagged $B^0$ plus $\bar{B^0}$ time 
 integrated event sample by doing a maximum likelihood fit to the 
 Probability Distribution Function (PDF) given by

\begin{equation}
PDF = \frac {  |\mathcal{A}^0|^2  
+ | \bar{\mathcal{A}}^0|^2  } {N^0 + \bar{N}^0} ,
\label{} 
\end{equation}

\noindent where $N^0 = \int |\mathcal{A}^0|^2 \, ds_{ij} \, ds_{jk}, 
$ and $\bar {N}^0 = \int |  \bar{\mathcal{A}}^0|^2 \, ds_{ij} \, ds_{jk}$ and 
$a, \delta$ and $\bar{a}, \bar{\delta}$  are free fit parameters.

In the standard single sample Dalitz plot amplitude analysis, described in
section II, two parameters, one magnitude and one phase, 
should be fixed. What is measured are the relative contributions and phase
differances. The magnitude sets the overall scale. This would be the case for
each of the two charged modes with a total of four fixed parameters.
In the joint fit, we have to fix three parameters, one magnitude, that sets the
overall scale for both modes, and two phases,
one for each c.c mode.
 In the joint
fit, the freedom of all magnitudes of one of the modes, say $\bar{B^0}$, is 
what guaranties that
we are able to have different total number of events for each mode.
In our particular case 
we fixed  $a_\chi, \delta_\chi$ and $\bar{\delta_\chi}$.
% and  $\bar{a_\chi}$ is kept free}. 
 We are able to measure a difference in number of events from 
$B^0$ and $\bar{B^0}$, searching for direct  global asymmetrie. With one single 
procedure it is 
possible to extract in an
independent way the parameters from the amplitudes $\mathcal{A}^0$ and 
$\bar{\mathcal{A}^0}$
and the ratio in number of events, calculated by the ratio of the 
normalizations $N^0/\bar{N}^0$.

\begin{figure}[hbt]
\centering
\includegraphics[scale=.35]{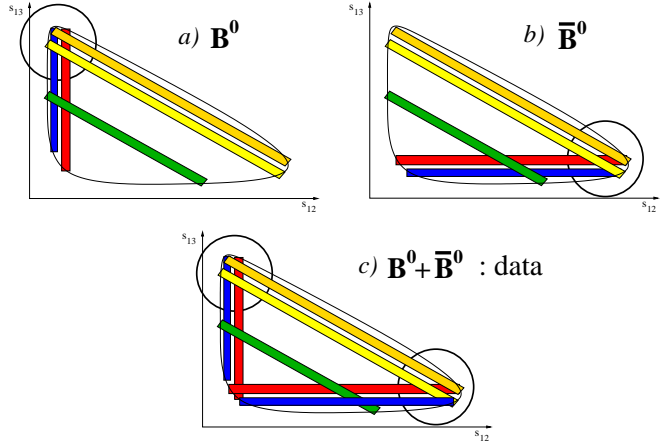}
\caption{Schematic representation for the Dalitz plot of $a) B^0 \to K_S\pi^+\pi^-$, 
$b) \bar{B^0} \to K_S\pi^+\pi^-$ and 
$c)$ the joint figure obtained for the untagged sample. Bands  pictorially
 represent resonances regions. The vertical and  horizontal  bands are $K\pi$
 resonances the in the diagonal are $\pi \pi$ overlapping resonances. For simplicity
we ignore angular effects in the distribution. Circles are displaying regions of
interference within each $B^0$ or  $\bar{B^0}$ decay. There is no interfence between the 
horizontal and the vertical bands for they do not belong to the same mode.  
It is the non-overlapping of circles that assures the separation between both amplitudes in the joint fit analysis.}
\label{fig:ampsdados}
\end{figure}

%explicar o porque podemos usar o ajuste misto
In the convention $(K_S, \pi^+, \pi^-)\to ({\bf 1,2,3})$ for the final state particle numbering,
the charge conjugation operation is equivalent to switching the Dalitz variables
 $s_{12} \leftrightarrow s_{13}$.
The resonances $K^*$ bands are aligned in different axis for 
$\mathcal{A}^0$ and $\bar{\mathcal{A}}^0$,
establishing some sort of signature for the event origin $B^0$ or $\bar{B^0}$, as can be seen schematically in Fig.
\ref{fig:ampsdados}. Although the $\pi \pi$ resonances for 
$B^0$ and $\bar{B^0}$ overlap,  
 the interference regions between $K\pi$ and $\pi\pi$
resonances are well separated. These interfence regions are vital for
establishing the phases, and they guarantee that the fitting procedure finds a unique
solution.
Although we cannot distinguish events, the joint fit can 
identify two different overlapping surfaces relative to $B^0$ and $\bar{B^0}$.

%{Como estava \bf  due to the non-overlapping of at least one
%interference region from both amplitudes. The non-overlapping
%assures that the amplitudes can be explored in an independent way by the fitting procedure,
%guaranteeing the unicity in the fit result.}
 %However, in the untagged analysis, data supplies a joint Dalitz plot, where a generic point
%have an unknown origin.
 %{A minha proposta :\bf one with interference comming 
%from a coherent sum and the other without interference, comming from an incoherent sum. These different
%overlapping behaviors  assures that the amplitudes can be explored in an independent way by the fitting procedure,
%guaranteeing the unicity in the fit result. }

%%%%%%%%%%%%%%%%%%%%%%%%%%%%%%%%%%%%%%%%%%%%%%%%%%%%%%%%%%%%%%%%%%%%%%%%%%%%%%%%%%%%%%%%%%%%%%%%%%%%%%%%%%%%%%%

\section{Feasibility Study}
To investigate the feasibility and the error size in the joint analysis,
we generated and fitted one hundred Monte Carlo samples of 
 100K $B^0$ plus $\bar B^0 $ events, that can be obtained  by  LHCb 
 experiment.  We do not consider any theoretical or
 experimental systematic uncertainties. Detector and background effects are as
 well not taken into account.

\begin{figure}[hbt]
\centering
\includegraphics{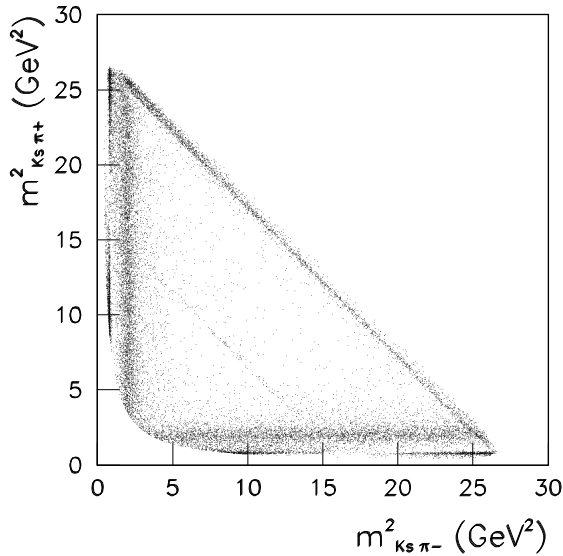}
\caption{$B^0$, $\bar B^0 \to K_s \pi^+ \pi^-$ events distribution from fast MC generated with table 
\ref{tab:viabilidade} parameters.}
\label{fig:dalitzviab}
\end{figure}

 The parameters  we used to generate the  Monte Carlo samples  are listed 
 in Table 
 \ref{tab:viabilidade}, column input.
They are close to the values obtained  in measured parameters for the 
    $B^{\pm} \to K^{\pm} \pi^+ \pi^-$ decays \cite{BaBark2pi, bellek2pi}. 
 For the     $B^0$ and  $\bar B^0 \to K_s \pi^+ \pi^-$ decays 
  we used the same 
     parameters as in the charged mode, except for  including a 
      {\em tree} contribution, discussed below, for  $K^*(890)$ and
     $K^*_{0}(1430)^0$  and the $\gamma$ phase,  taken as $69^o$. 
 The Dalitz plot distribution for one generated experiment of 100K events is displayed in
 Fig.  
\ref{fig:dalitzviab}, where the small contribution of the resonance $\chi_{c0}$ can be seen.

One important issue concerning the $\gamma$ extraction, is the size of 
the ratio $r=a_T/a_P$ and the phase
difference $\theta=\delta_T - \delta_P$ from the {\em tree} and 
{\em penguin} amplitudes of the $K^*$ resonances.
 The smallest the value of $r$ the more difficult it is to apply
this procedure to measure $\gamma$.
The theoretical knowledge of these quantities is model dependent.
Some groups use factorization approach \cite{teo-beneke}, and obtain large $r$ and small $\theta$.
On the other hand, non-factorisable approach for pseudoscalar-pseudoscalar $B$ decay \cite{buras},
presents an opposite scenario, with small $r$ and large $\theta$. The joint fit, applied to the
real data, will be able to define which theoretical approach is more adequate, 
since it is possible to measure the parameters under discussion. 
For our study, we chose $r=0.45$.

\begin{table}[htb]
\begin{center}
  \begin{tabular}{c|c|c|c}
    
decay &$B^0/\bar B^0$ & input &  fit 100K events  \\
    \hline\hline
$\chi_c K_s$ & $a_0/\bar a_0$ & 0.30/0.30 &  fixed/$(0.30 \pm 0.03)$  \\
& $\delta_0/\bar\delta_0 $ & 3.78/3.78  & fixed/fixed \\
 \hline
$K^*(890)\pi$ & $a_1/\bar a_1$ & 1.17/1.30 & $1.17 \pm 0.06 $  /$(1.30 \pm 0.01)$ \\
 & $\delta_1/\bar \delta_1$ & 0.40/5.98 &  $(0.41 \pm 0.08)$/$(5.99 \pm 0.07)$  \\
 \hline
$K_0^{*}(1430)\pi$ & $a_2/\bar a_2$ & 2.45/2.72 &  $(2.45 \pm 0.11)$/$(2.72 \pm 0.13)$  \\
 & $\delta_2/\bar \delta_2$ & 0.375/6.00 &   $(0.39 \pm 0.08)$/$(6.00 \pm 0.06)$ \\
 \hline
$\rho^0 K_s$ & $a_3/\bar a_3$ & 0.60/0.60 &   $(0.60 \pm 0.04)$/$(0.60 \pm 0.04)$ \\
& $\delta_3/\bar \delta_3$ & 1.20/1.20  &  $(1.22 \pm 0.09)$/$(1.20 \pm 0.07)$ \\
 \hline
$f_0 K_s$ & $a_4/\bar a_4 $ & 1.03/1.03 &  $(1.02 \pm 0.06)$/$(1.04 \pm 0.05)$  \\
& $\delta_4/\bar \delta_4$ & 2.30/2.30 &   $(2.30 \pm 0.07)$/$(2.30 \pm 0.08)$ \\
 \hline
 \end{tabular}
  \begin{tabular}{c}
  $N^0( B^0 \to K_s \pi^+ \pi^-)/ N^0(\bar B^0 \to K_s \pi^+ \pi^-)$  =
  0.84 $\pm$  0.12  \\
 \hline\hline
  \end{tabular}
  \caption[]{ Monte Carlo simulation for $B^0$,   $\bar B^0 \to K_s \pi^+ \pi^-$ 
  decay. We generated sample with the the parameters $a_i$ and $\delta_i$ for 
  $B^0$ and $\bar a_i$ and $\bar \delta_i$ for $\bar B^0 $. The third 
  column display the fit results with one hundred samples for each  
  100K events. The last line is the ratio between $B^0, \bar B^0 \to K_s \pi^+ \pi^-$
  number of events.}
  \label{tab:viabilidade}
\end{center}
\end{table}

As stated in section IV, the first step of the method would be to solve the 
charged mode Dalitz Plot, from which hypothesis 1 could be tested. The parameters $a_P$ and 
$\delta_P$  would then be input to the neutral system fit. 
Since charged mode analysis presents no novelty and the statistical errors are expected to
be  low compared to the neutral decays we skip this step for this study.

Each 100K event sample of  $B^0$ plus $\bar{B^0}$ generated with the amplitudes and
phases listed in the column input of Table \ref{tab:viabilidade} were fitted 
with an
unconstrained model with just three parameters fixed. In the last column of 
Table \ref{tab:viabilidade} we display the central values
 of the Gaussian distributions of the 100 experiment's fit result and 
the errors are the  width of the same Gaussian distribution. The total number of events of
each  $B^0$ and $\bar{B^0}$ is the result of the integration of either amplitude within
the Dalitz plot.
The fact that the $\chi_{c0}$ magnitude for  $\bar{B^0}$ is measured to be equal to the fixed
$a_0$ is the test for hypothesis 3.
 The extracted quantities are in agreement
with the generated ones and with small errors, stating the feasibility of the joint fit method.
Using the parameters from Table \ref{tab:viabilidade} and the relations
developed in section IV, we measure $\gamma= 69^\circ \pm 7^\circ$.

If in the unconstrained analysis, the three hypotheses prove to be correct, we
can reduce the numbers of free parameters assuming that the $K^*$
resonances parameters from $B^0$ and $\bar{B^0}$ are given by the following {\em tree} and {\em penguin} sum:
\begin{eqnarray}
a \, e^{i \, \delta}= a_P \, e^{i \, \delta_P} + a_T \, e^{i \, \delta_T +\gamma}, \nonumber \\
\bar{a} \, e^{i \, \bar{\delta}}= a_P \, e^{i \, \delta_P} + a_T \, e^{i \,
\delta_T -\gamma}.
\label{eq:pingtreekstar}
\end{eqnarray}

Instead of fitting for $a, \delta, \bar{a}$ and $ \bar{\delta}$, we  fit for $a_T,
\delta_T  $ and $\gamma$ since we can fix $a_P, \delta_P$ from 
$B^{\pm} \to K^{\pm} \pi^+ \pi^-$ result.
This way, constraining the fit with the hypothesis previously tested we reduce
the error in $\gamma$. We measure with the constrained hypothesis, 
$\gamma= 69^\circ \pm 5^\circ$ and also we obtained for the ratio between 
the {\em tree} and {\em penguin} component $ r= 0.45 \pm 0.05$ 
.

%%%%%%%%%%%%%%%%%%%%%%%%%%%%%%%%%%%%%%%%%%%%%%%%%%%%%%%%%%%%%%%%%%%%%%%%%%%%%%%%%%%%%%%%%%%%%%%%%%%%%%%%%%%%%%%

\section{Conclusion}

We discussed some aspects of the search of \acp in three body B decays,
emphasising  the main characteristic of these decays, which is the interference 
among two neighbor resonances in the Dalitz plot. In particular,  
we discuss how  the CKM  $\gamma$ phase can be  extracted applying  the 
Isobar model\cite{pdg},  extensively used in   many  charm meson three 
body decay amplitude analysis. We showed that the phase differences, 
coming from  the amplitude analysis, can provide  
richer  information  than treating the  three body decays as a 
straightforward extension  of  two body hadronic decays. 
 We showed that if there is a weak phase difference between two intermediate  state,
\acp will be present in the Dalitz plot regardless of the strong phase difference.

We presented  a method to extract the CKM $\gamma$  angle using a 
combined Dalitz plot analysis from $B^{\pm} \to K^{\pm} \pi^+ \pi^-$ 
and  $B^0, \bar B^0 \to K_s \pi^+ \pi^-$. 
This method use three basic hypothesis that can be tested before one proceeds to the constrained fit.
For measuring the $B$ neutral analysis, we use a new technique of joint fit
that allows us to extract, in an independent way, the amplitudes from two summed surfaces in 
a joint sample of $B^0 + \bar B^0$ untagged events. 
We carried a simplified fast MC study to estimate the errors associated to the joint fit technique. 
Assuming a relatively low
statistics for the anchor resonance $\chi_{c0}$ we 
obtained $\gamma$ with a $5^\circ$ error.

During the analysis we can measure CP violation by counting the number of events
 from $B^0$ and $\bar{B^0}$,
information extracted in the joint fit,
or exploring Dalitz symmetries as discussed in \cite{burdman}.
We can also measure the ratio and phase difference from tree and {\em penguin} 
amplitudes of the $K^*$ resonance, 
defining which theoretical approach, factorisable or non-factorisable, 
 is more adequate.

The method presented here is competitive with the 
other approaches to determine the CKM  $\gamma$ angle \cite{ schneider}
and needs   the high  statistics expected for the LHCb experiment, 
due to the  small contribution of the reference channel $B\to \chi^0 K$. 
However, in the case  that the  $ f_0(980) $ resonance is dominated by the $ s \bar s$ component
\cite{bnn},   or even if the ratio between the tree and {\em penguin} is  
negligible \cite{f0},  the $B\to f_0(980) K$ amplitude could take the place 
of the charmonium as a reference channel in the analysis. 

\vskip 0.3cm
\noindent{\bf Acknowledgements}
\vskip 0.2cm

{We thank  Professor Giovanni  Bonvicini for  suggestions and 
comments.}

\end{document}